\documentclass[superscriptaddress,twocolumn,10pt]{revtex4-1}

\usepackage{amsmath,graphicx,color}
\usepackage{booktabs,array,dcolumn}
\usepackage{multirow}
\usepackage{times}
\newcolumntype{C}{>{\centering}p}

\begin{document}

\title{Polytypism in the ground state structure of the Lennard-Jonesium}

\author{L\'\i via B. P\'artay}
\affiliation{Department of Chemistry, University of Reading, Reading RG6 6AD, UK}
\author{Christoph Ortner}
\affiliation{Mathematics Institute, University of Warwick, Coventry CV4 7AL, UK }
\author{Albert P. Bart\'ok}
\affiliation{STFC Scientific Computing Department, Rutherford Appleton Laboratory, Didcot OX11 0QX, UK}
\author{Chris J. Pickard}
\affiliation{Department of Materials Science \& Metallurgy, University of Cambridge, Cambridge CB3 0FS, UK}
\affiliation{Advanced Institute for Materials Research, Tohoku University 2-1-1 Katahira, Aoba, Sendai, 980-8577, Japan}
\author{G\'abor Cs\'anyi}
\affiliation{Engineering Laboratory, University of Cambridge, Cambridge CB2 1PZ, UK}

\date{\today}

\begin{abstract}
We present a systematic study of the stability of nineteen different periodic structures using the finite range  Lennard-Jones potential model discussing the effects of pressure, potential truncation, cutoff distance and Lennard-Jones exponents.
The structures considered are the hexagonal close packed (hcp), face centred cubic (fcc) and seventeen other polytype stacking sequences, such as dhcp and $9R$.
We found that at certain pressure and cutoff distance values, neither fcc nor hcp is the ground state structure as previously documented, but different polytypic sequences.
This behaviour shows a strong dependence on the way the tail of the potential is truncated.
\end{abstract}

\maketitle

%%%%%%%%%%%%%%%%%%%%%%%%%%%%%%%%%%%%%%%%%%%%%%%%%%%%%%%
%%%
%%%   introduction
%%%

\section{Introduction}

Polytypism is a special form of polymorphism, occurring in layered materials, in which the polymorphs are derived simply by varying the way in which the layers are arranged relative to each other.
This means that the various stacking arrangements do not affect the chemistry of the phase as a whole, but some of the physical properties (e.g. density, Young modulus, band gap or electron mobility) can be significantly different.
A large variety of materials have several different stable polytype phases~\cite{poly_review}, one of the most extensively studied being SiC~\cite{Jepps,SiC_raman}.
SiC has more than 200 identified polytypes, a few being more favoured in applications than the rest due to their superior electronic properties.
Many materials with similar structural properties also form polytypes, such as metal sulphides and halogenides, e.g. ZnS~\cite{ZnS,ZnS_theory} and CdI$_2$~\cite{CdI2}.
The physical properties of such materials can be tuned by changing the stacking sequences, e.g. in ZnO~\cite{ZnO}.
Polytypism also occurs in the case of diamond. The common cubic form of diamond has a hexagonal polytype called Lonsdaleite, which is suggested to be a complex mixture of different stacking sequences \cite{diamond,diamond_xray}. Similarly, hexagonal (Ih) and cubic (Ic) ice are also polytypic structures.
Some elements are also known to form polytype phases, such as lanthanum, which exists in the dhcp form~\cite{La_dhcp}, samarium and lithium having the $9R$ stacking sequence as the ground state structure~\cite{Sm_9R,Li_9R}, erbium which is stable in both dhcp and 9R stacking sequences at different pressures~\cite{Er_dhcp_9R}, and bismuth, long suspected to exist in several polytypic forms~\cite{Bi_poly}.
It has been speculated that the transformation from fcc to hcp structure with increasing pressure might occur through a series of different stacking fault structures, e.g. as in the case of noble gases xenon and krypton,  suggested by some experimental results~\cite{Jephcoat_KrXe,Errandonea:2002hx}, or in the case of iron at high pressure and high temperature~\cite{dhcp_Fe1,dhcp_Fe2}.
Finally, if a wider definition of polytypism is used such that structurally compatible modules are also considered, a range of minerals which include the pyroxenes, perovskites, spinelloids, chlorites and oxides form polytypic structures as well~\cite{Price:1984vo}.

In order to model the polytypic behaviour, the axial next-nearest-neighbour Ising (ANNNI) model, was used in the 1980s~\cite{Price:1984vo, YEOMANS_Price1,YEOMANS:1988tl}.
(An overview of the ANNNI and A3NNI ground state structures are given in the Appendix.)
However, the two possible layer types in ANNNI, usually marked by $\downarrow$ and $\uparrow$ are interchangeable, a phase is only defined by the number of consecutive layers of the same orientation, but not the orientation of the layer itself, thus phases $\downarrow\downarrow\uparrow$ and $\uparrow\uparrow\downarrow$ are identical.
In contrast, close packed stacking structures are built up by layers in three different possible positions, usually denoted by A, B and C, forming either a hexagonal (in ABA stacking) or a cubic (in ABC stacking) layer. These two are not identical nor interchangeable, meaning that the ability of ANNNI to describe the behaviour of close packed materials is limited.

One of the most widely used models to study close-packed materials is the Lennard-Jones pair potential.
It has been long known that its low temperature dominant structures are the hexagonal close packed (hcp) and the face centred cubic (fcc)~\cite{Koba_hcpfcc_52}.
Interestingly, although other structures (bcc, simple cubic, diamond) have been studied~\cite{DLT_LJ_crystal}, to the best of our knowledge no other polytype sequences have ever been investigated from the point of view of phase  stability, only as stacking faults in relation to crystal growth defects or nucleation~\cite{Wales_LJ2016}.
It is also notable that the customary finite range  truncation of the potential  has a significant effect on the liquid-vapour equilibrium~\cite{BSmit_LJ_vap,Panagiotopoulos_LJ_vap,Johnson_LJ_vap} and on the melting temperature~\cite{dePablo_LJ_melt,Sadusa_LJ_melt},
yet it is rarely mentioned and almost never taken into account in the discussion of the low temperature solid phases, causing an apparent inconsistency in the literature with regard to  the  lowest energy  structure: some works refer to the fcc~\cite{Wales_LJ2016,LJ_fcc_Hoef}, others to hcp~\cite{Waal_LJ,DLT_LJ_crystal} as the global minimum of the Lennard-Jones model.
An exception is an article by Jackson et. al~\cite{LJ_cutoffs_Ackland} showing that the ground state can be either fcc or hcp depending on the cutoff distance and method of truncation.

Our aim in this work is to provide a systematic study of the ground state structure of the Lennard-Jones potential  considering different polytypic stacking sequences, and fill the gap in the literature regarding its dependence on pressure, potential truncation and potential parameters.
The rich diversity of structures we find serve as a reminder that complex material behaviour can result from comparatively simple models, and that implementation details can have a strong effect on phase stability when modelling materials.

\section{Computational details}

A generalised form of the Lennard-Jones  potential can be given by
\begin{equation}
U_{\mathrm{LJ}}(r) = \frac{p}{(p-q)(\frac{q}{p})^{q/(p-q)}} \epsilon \bigg[ \Big(\frac{\sigma}{r}\Big)^{p} - \Big(\frac{\sigma}{r}\Big)^{q} \bigg].
\end{equation}
The values $p=12$ and $q=6$ are most common, in which case one obtains
\begin{equation}
U_{\mathrm{LJ}}(r) = 4\epsilon \bigg[ \Big(\frac{\sigma}{r}\Big)^{12} - \Big(\frac{\sigma}{r}\Big)^6 \bigg],
\end{equation}
where $\epsilon$ is the depth of the potential well, $\sigma$ is the size of the repulsive core  and $r$ is the distance between two particles.
This potential is usually truncated at a cut-off distance, $r_c$, and to avoid the discontinuity at this point, the potential can be shifted.
The {\it energy-shifted} LJ potential is a continuous ($C^0$) function,
\begin{equation}
U_{\mathrm{LJ}-C^0}(r)=
\begin{cases}
     U_{\mathrm{LJ}}(r) - U_{\mathrm{LJ}}(r_c)    & r \leq r_c \\
    0          & r > r_c. \\
  \end{cases}
  \label{eq:lj_e_shift}
\end{equation}
In order to obtain continuous forces at the cut-off distance, and thus make the potential function differentiable, it can be {\it force-shifted}, which leads to
\begin{equation}
U_{\mathrm{LJ}-C^1}(r)=
\begin{cases}
    \!\begin{aligned}
     U_{\mathrm{LJ}}(r) & - U_{\mathrm{LJ}}(r_c) - \\
     & -(r-r_c)U'_{\mathrm{LJ}}(r_c)
     \end{aligned} & r \leq r_c \\
     0          & r > r_c. \\
\end{cases}
\end{equation}
In order to make also the second derivatives continuous at $r_c$ (i.e. create a $C^2$ function), the potential can be further shifted by a third term,
\begin{equation}
U_{\mathrm{LJ}-C^2}(r)=
\begin{cases}
     \!\begin{aligned}
     U_{\mathrm{LJ}}(r) - U_{\mathrm{LJ}}(r_c) &\\
      - (r-r_c)U'_{\mathrm{LJ}}(r_c) & \\
      - {\textstyle \frac{1}{2}}(r-r_c)^2 U''_{\mathrm{LJ}}(r_c) &
     \end{aligned} & r \leq r_c \\
    0          & r > r_c. \\
  \end{cases}
  \label{eq:c2_shift}
\end{equation}
Alternatively, a sigmoidal shaped function, $f_s(r)$, can be used to "smooth out" the force shifted potential within a distance $r_{s}$ of the cutoff,
\begin{equation}
U_{\mathrm{LJ}-C^1-\mathrm{smoothed}}(r)= f_s(r)  U_{\mathrm{LJ}-C^1}(r) .
\end{equation}
%For $f_s(r)$ we used two different functions, a $C^3$ function, a polynomial in the form (where $x=[r-(r_c-r_s)]/r_s$)
%\begin{equation}
%%%this function is the "smootheststep" derived by Kyle McDonald and first posted to Twitter with a derivation on GitHub.
%f_s(r)
%\begin{cases}
%     1  & r \leq (r_c-r_s) \\
%     \!\begin{aligned}
%     1-35 x^4&+84x^5 - \\
%     & -70x^6+20x^7
%      \end{aligned} & (r_c-r_s) < r \leq r_c \\
%     0   & r > r_c, \\
%  \end{cases}
%\label{eq:c3_poly}
%\end{equation}
For $f_s(r)$ we used an infinitely differentiable ($C^\infty$) function,
\begin{equation}
f_s(r)
  \begin{cases}
     1  & r \leq (r_c-r_s) \\
     1-\frac{e^{(-1/x)}}{e^{(-1/x)}+e^{(-1/(1-x))}} & (r_c-r_s) < r \leq r_c \\
     0   & r > r_c, \\
  \end{cases}
  \label{eq:cinf_exp}
\end{equation}
where $x=[r-(r_c-r_s)]/r_s$).

%\lbp{C3 smoothing is commented out as we agreed that it does not give additional info.}

\begin{figure}[hbt]
\begin{center}
\includegraphics[width=6.5cm,angle=90]{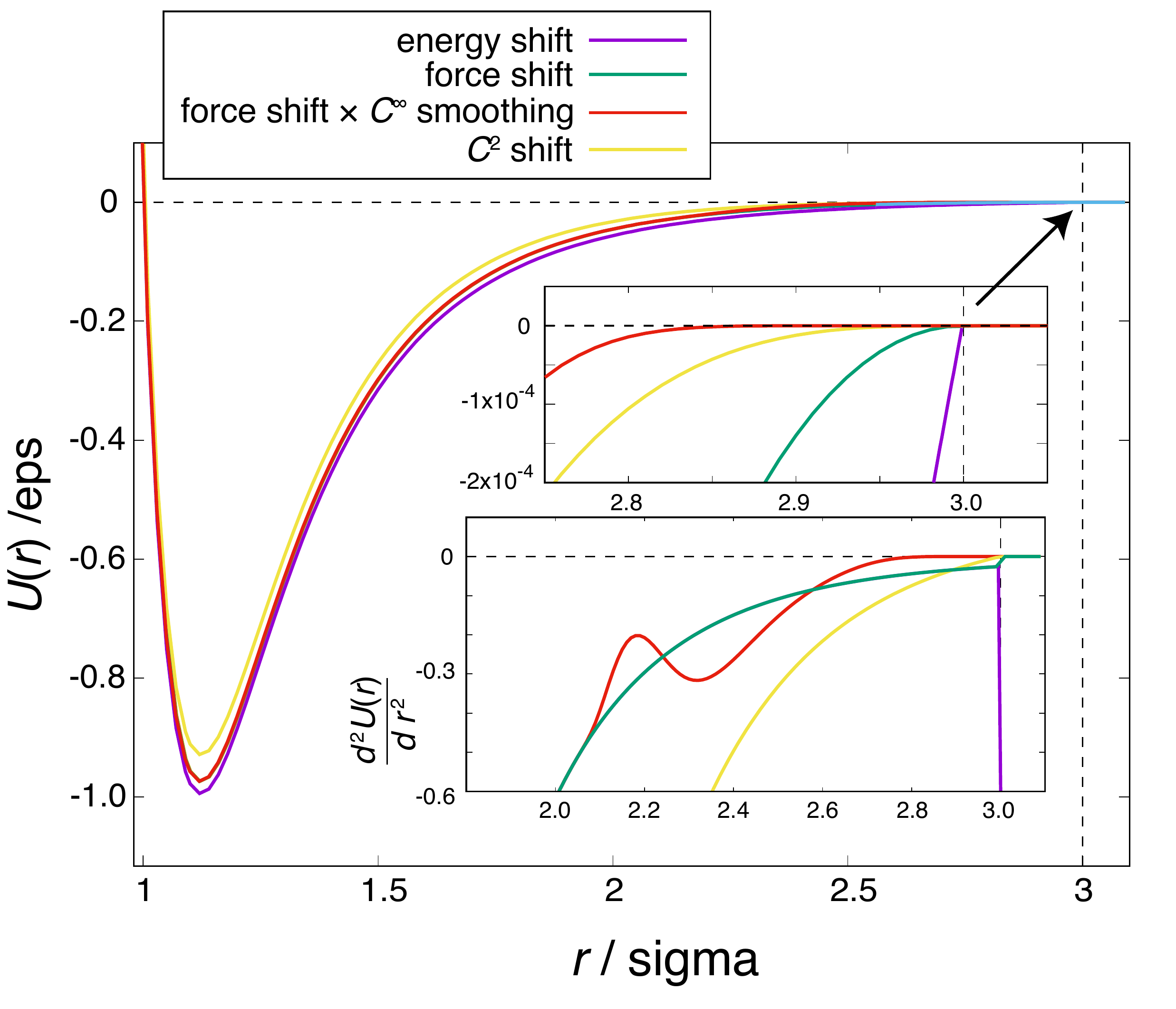}
\end{center}
\vspace{-10pt}
\caption {Shape of the Lennard-Jones potential of exponents $p=12$ and $q=6$, with different cutoff schemes.
All functions were used with cutoff distance $r_c=3.0 \sigma$ and in case of the smoothed force-shifted potential, $r_s=1.0 \sigma$ smoothing range has been used.
The top inset shows the functions at the vicinity of the cutoff, while the bottom inset shows the second derivative at the potentials.}
\label{fig:LJ_r_cutoff_types}
\end{figure}

\begin{table}[h]
\caption{Different stacking variants studied, listed with both their hc and ABC notation.
Alternative names and Ramsdell notations are shown for specific stacking sequences in parenthesis.}
\begin{tabular}{p{0.78in}cll}
\hline\hline
stacking & min. number & physical stacking of the layers \\
variants & of layers        &   \\  %
\hline
c (fcc, $3C$)          & 3  &  $[ABC]_n$  \\  %
h (hcp, $2H$)          & 2  &  $[AB]_n$     \\
hc (dhcp,$4H$)      & 4  &  $[ABCB]_n$        \\
hcc (thcp,$6H_1$)      & 6  &  $[ABCACB]_n$         \\
hccc           & 8  &  $[ABCABACB]_n$          \\
hcccc          & 10 & $[ABCABCBACB]_n$        \\
hccccc         & 12 &  $[ABCABCACBACB]_n$         \\
hhc ($9R$)    & 9  &  $[ABACACBCB]_n$         \\
hhcc           & 12  &  $[ABACBCBACACB]_n$          \\
hhccc          & 5 & $[ABACB]_n$        \\
hhcccc        & 18 &  $[ABACBACACBACBCBACB]_n$        \\
hhhc           & 8 & $[ABABCBCB]_n$        \\
hhhcc          & 10 & $[ABABCACACB]_n$        \\
hhhccc         & 12 & $[ABABCABABACB]_n$        \\
hhhhc         & 15 & $[ABABACACACBCBCB]_n$        \\
hhhhcc         & 18 & $[ABABACBCBCBACACACB]_n$        \\
hhhhhc        & 12 & $[ABABABCBCBCB]_n$        \\
hchcc ($15R$)  & 15 &  $[ABCBACABACBCACB]_n$         \\
hchhc           & 10 &  $[ABCBCACBCB]_n$         \\
\hline\hline
\end{tabular}
\label{table:stacking_var}
\end{table}

\begin{figure}[hbt]
\begin{center}
\includegraphics[width=3cm]{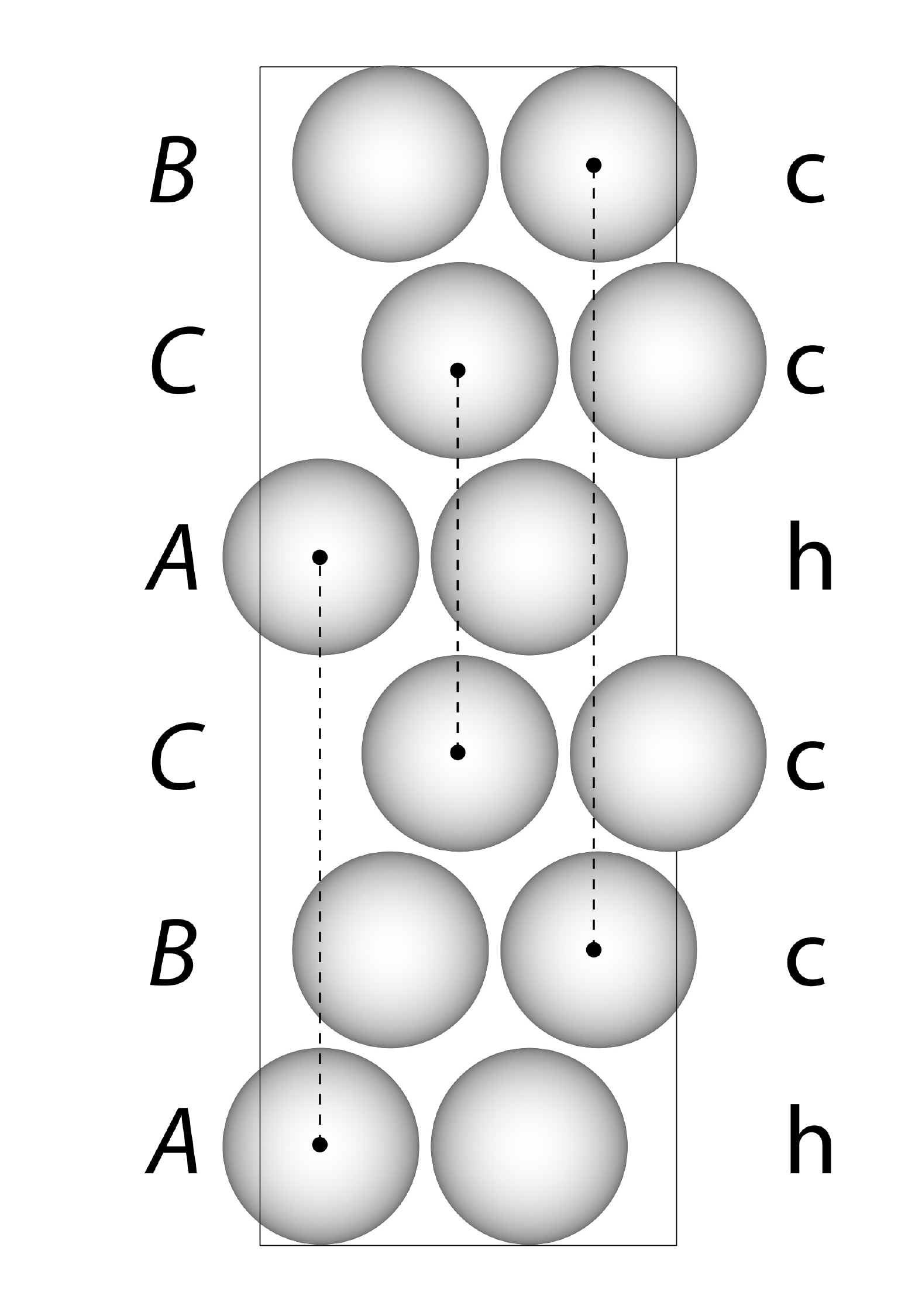}
\end{center}
\vspace{-20pt}
\caption {An example snapshot\cite{Li:2003wy} of the $\langle \mathrm{hcc} \rangle$ sequence. Layers at the same lateral position are connected by dashed lines.}
\label{fig:HFF_example}
\end{figure}

Nineteen different stacking sequences were considered (shown in Table~\ref{table:stacking_var}), all variations up to five stacking layers, and some of the possible six layer arrangements.
The notation used to mark the sequences is that layers with fcc surroundings (the two neighbouring layers occupy different positions) is marked ``c'' as cubic, while the layers with hcp surrounding (sandwiched between two layers occupying the same position) is marked ``h'' as hexagonal.
An example structure, with associated notation, is shown in Figure~\ref{fig:HFF_example}.

The geometry optimisations were performed with the QUIP package~\cite{quip}, with the conjugate gradient method and double-checked with the steepest descent method for several cases.
The minimisation tolerance was set to $10^{-6} \epsilon/\sigma$ for the norm of the forces, which corresponds to $10^{-15} \epsilon/\mathrm{atom}$ accuracy in the energy calculation. Hydrostatic pressure was applied.
The minimisations were started from configurations where atoms were placed $1.0\sigma$ distance from each other, and during the minimisation the atomic positions and all the lattice parameters were allowed to relax.
The calculations were done with different shifted and smoothed versions of the potential to compare their effects within the truncation range $2.0\sigma \leq r_c \leq 6.0\sigma$, in $0.05\sigma$ intervals.

\section{Results: ground state phase diagrams of the Lennard-Jones potential}

\begin{figure}[htb]
\vspace{-10pt}
\begin{center}
\includegraphics[width=6cm, angle=90]{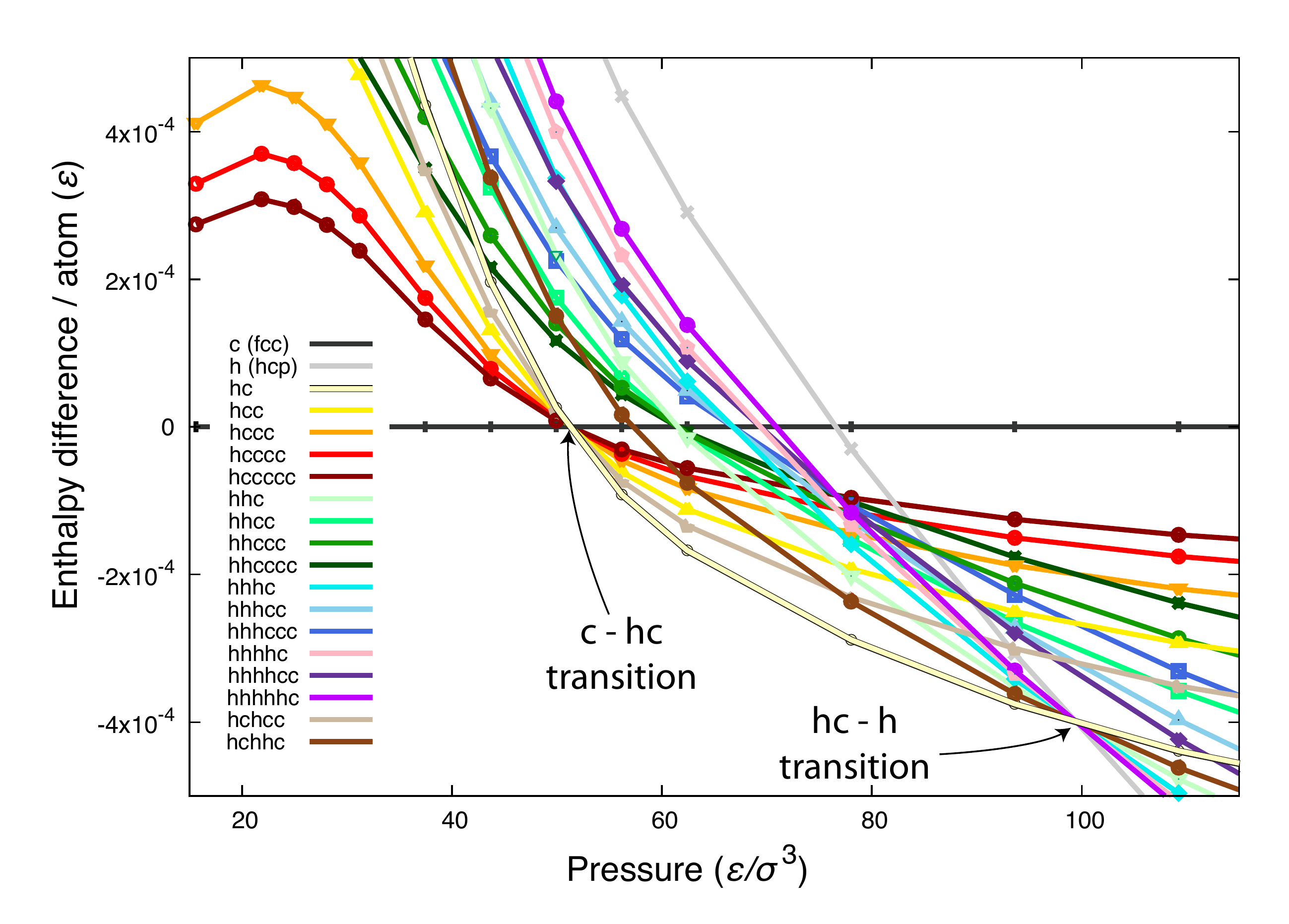}
\end{center}
\vspace{-20pt}
\caption {Enthalpy difference of different polytypes relative to that of the fcc structure as a function of pressure.
The potential is force shifted and cutoff is $r_{cut}=2.7\sigma$.
The arrows show the location of the two phase transitions where the enthalpy of several polytypes are almost equal due to the relatively short cutoff, thus the phase boundary between $\langle \mathrm{c} \rangle$ and $\langle \mathrm{hc} \rangle$ is close to being a multiphase boundary with $\langle \mathrm{hc}^k \rangle$ and $\langle \mathrm{hchcc} \rangle$ type polytypes, while the boundary between $\langle \mathrm{h} \rangle$ and $\langle \mathrm{hc} \rangle$ is degenerate with $\langle \mathrm{h}^k\mathrm{c} \rangle$ and $\langle \mathrm{hchhc} \rangle$  type sequences.
This behaviour becomes less prominent as the cutoff increases.}
\label{fig:LJ_enthalpy_diff}
\end{figure}

During the minimisation process configurations retained their stacking order, and the atoms forming the stacking plane also stayed perfectly in the plane.
Lattices remained orthorhombic, but the lattice height corresponding to the stacking direction changed with respect to the other two, as expected.
To be able to identify the ground state structures and draw the phase diagrams, the enthalpies of the minimised configurations were calculated and compared.
These enthalpy curves were individually checked and more calculations were performed with a finer pressure scale whenever it was necessary, thus we believe that no phases have been missed.
A set of example enthalpy curves can be seen on Figure~\ref{fig:LJ_enthalpy_diff}.

\begin{figure}[hbt]
\begin{center}
\includegraphics[width=8.5cm, angle=0]{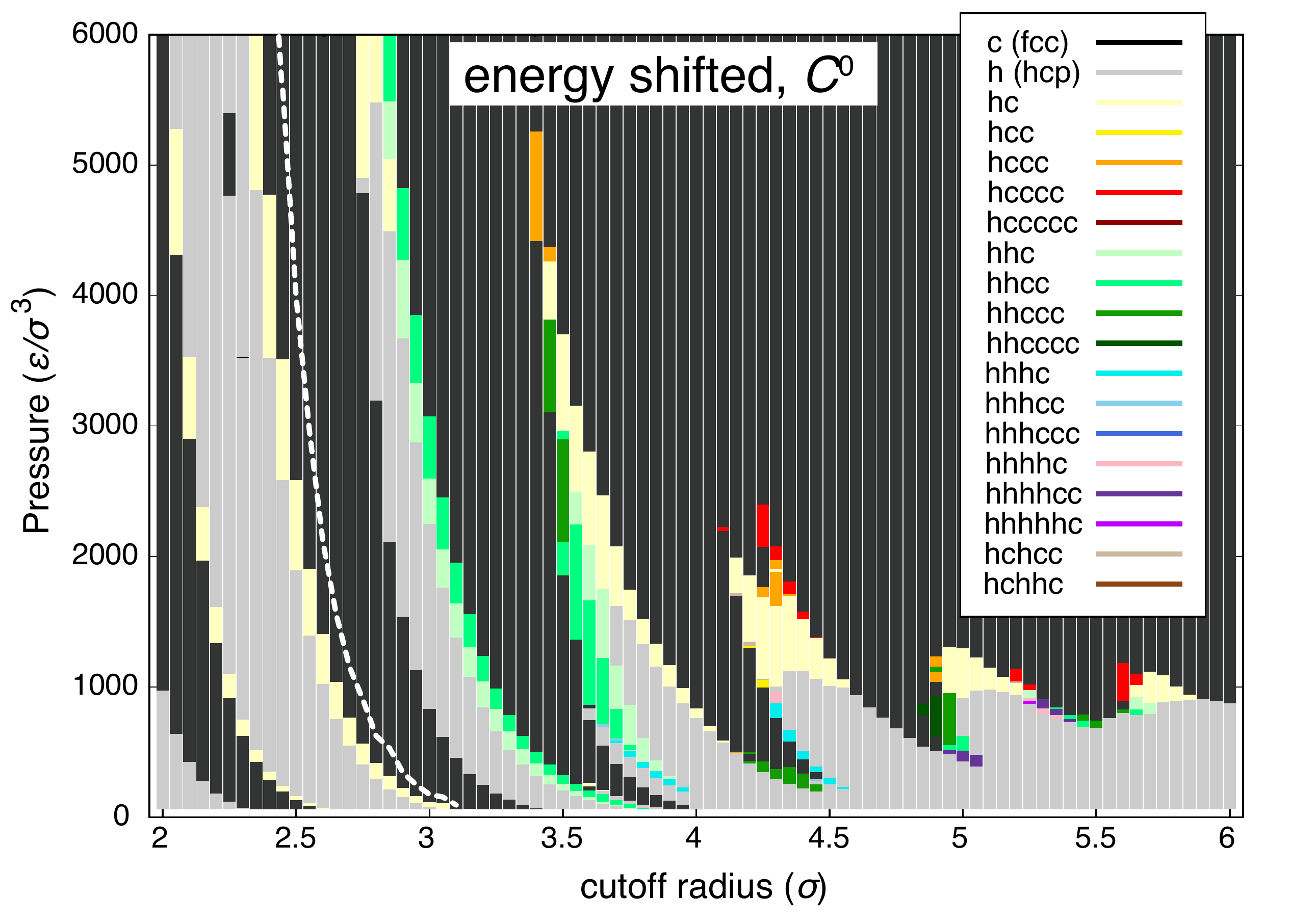}
\end{center}
\vspace{-20pt}
\caption {Lennard-Jones ground state structures as a function of cut-off radius and pressure.
The potential is $C^0$, i.e. energy shifted.
The colours represent the different stacking sequences, dark grey is pure fcc and light grey is pure hcp.
The white dashed line demonstrates one of the boundaries along which a new shell of atoms gets inside the cutoff sphere.}
\label{fig:LJ_pd}
\end{figure}

\begin{figure}[hbt]
\begin{center}
\includegraphics[width=8.5cm]{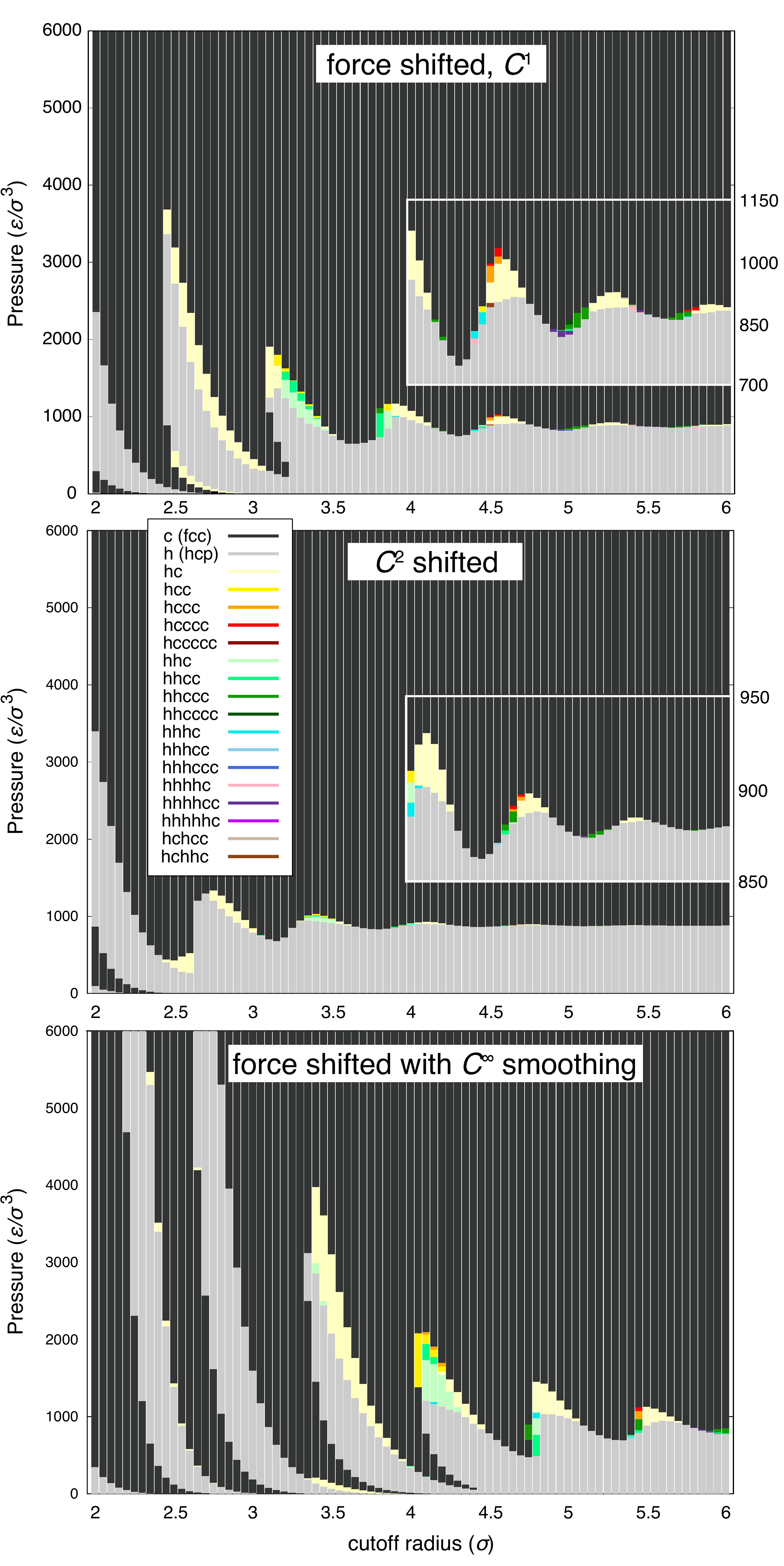}
\end{center}
\vspace{-20pt}
\caption {Lennard-Jones ground state structures as a function of cut-off radius and pressure.
In the top panel the potential is force shifted, the middle panel shows the $C^2$ shifted potential (see eq.~\ref{eq:c2_shift}), while the bottom panel shows the force shifted potential with a $C^\infty$ smoothing function (see e.q.~\ref{eq:cinf_exp}) applied in the $1.0 \sigma$ range of the cutoff.
The colours represent the different stacking sequences.
The insets show the phase boundaries between cutoffs $4.0\sigma$ and $6.0\sigma$ enlarged, the different pressure scale shown on the right.}
\label{fig:LJ_Fshift_pd}
\end{figure}

\begin{figure*}[hbt]
\begin{center}
\includegraphics[width=17.5cm]{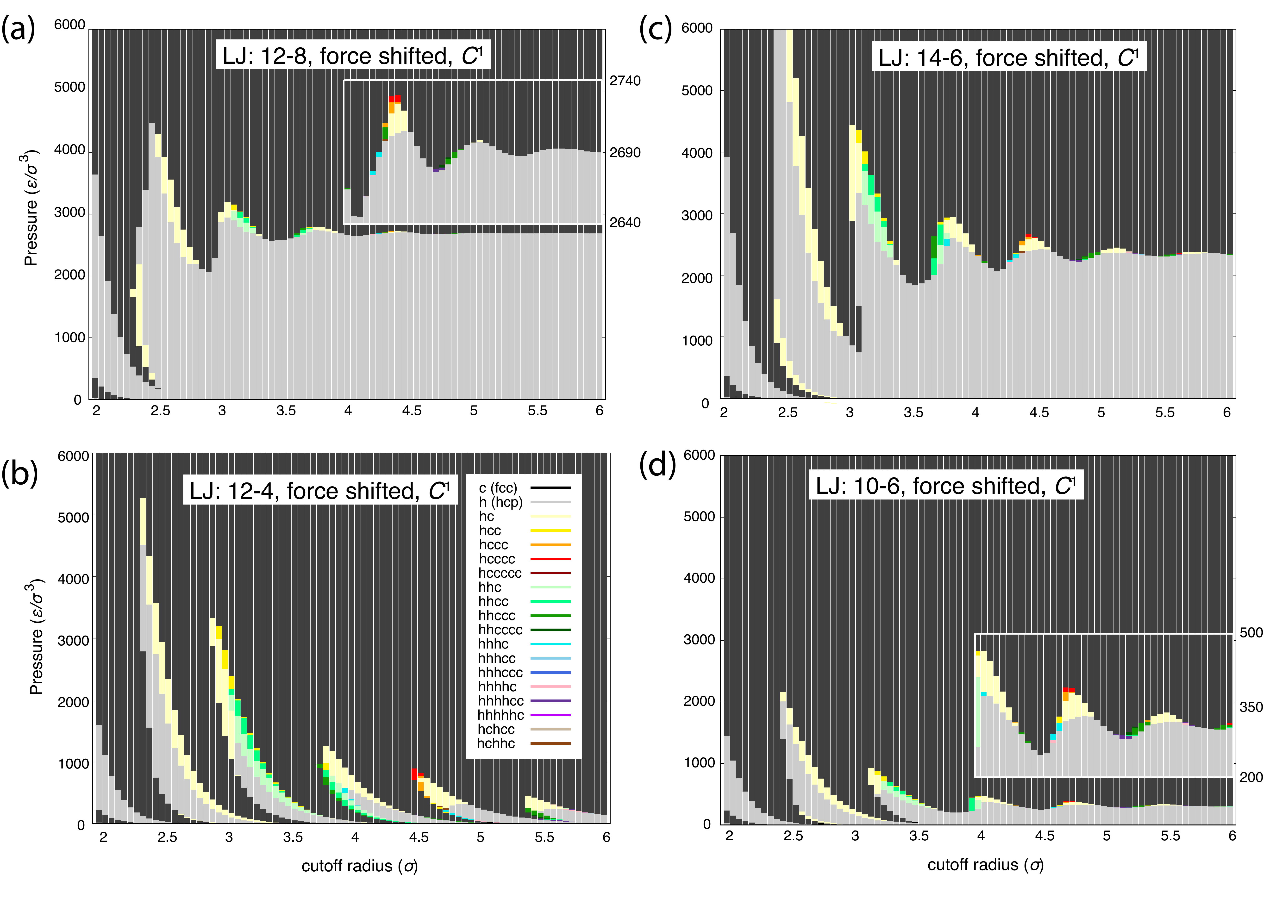}
\end{center}
\vspace{-30pt}
\caption {Ground state structure of Lennard-Jones potential with different exponents (a) $p=12$ and $q=8$, (b) $p=12$ and $q=4$, (c) $p=14$ and $q=6$, (d) $p=10$ and $q=6$ as a function of cut-off radius (in sigma units) and pressure.
The potential is force shifted.
The colours represent the different stacking sequences, dark grey is pure fcc and light grey is pure hcp.
The insets show the phase boundaries between cutoffs $4.0\sigma$ and $6.0\sigma$ enlarged, the different pressure scale shown on the right.}
\label{fig:LJ_X_6_pd}
\end{figure*}

Truncation distance vs. pressure phase diagrams of the Lennard-Jones type potentials with different cutoff schemes are shown on Figures~\ref{fig:LJ_pd}, \ref{fig:LJ_Fshift_pd} and \ref{fig:LJ_X_6_pd}.
The coloured regions show the series of phases found to be the most stable at a given truncation length and in a given pressure range.
Dark grey colour corresponds to the fcc and light grey to hcp structures, with other colours representing different stacking variants (yellow and red shades represent stackings with a single ``h" layer in the repeated subunit, green shades are polytypes with two consecutive ``h" layers, while blues and purples correspond to three and four consecutive ``h" layers, respectively).

It is clear from all the phase diagrams, that the hcp structure tends to be the most favourable stacking variant at lower pressure values, while for every value of the cutoff there is a pressure above which the fcc is the most stable polytype.
In order to see whether one of the studied polytypes becomes the ground state again at even higher pressures, the structures were minimised up to $p=6\times10^{5} \epsilon/\sigma^3$ for a few randomly chosen cutoffs: fcc remained the lowest enthalpy structure in every case.

However, the most striking result is that at the boundary between the ground state regions of fcc and hcp structures, several other stacking variants are found to be more stable.
This means that, in contrast with the common belief, the (truncated) Lennard-Jones potential exhibits a wide range of different global minima depending on the fine details of the potential.

The boundary between the fcc and hcp regions appear to have a ``wave''-like pattern for all the potential function variants we used.
The shape of these waves reflect how the distance of atomic shells decreases as the density increases with increasing pressure.
(The white dashed graph in Figure~\ref{fig:LJ_pd} represents the curve along which the number of atoms within the cutoff radius jumps from 177 to 201 in the fcc crystal.)
As different polytypes have different numbers of neighbours in each shell, their relative energy will be different depending on which shells lie within the cutoff radius.
As the pressure increases the atoms get closer, farther shells appear within the smaller cutoff radii, causing the phase boundary to be shifted towards smaller cutoffs.

At small cutoff values, fewer polytypes appear along the fcc-hcp boundary and these remain the same as the pressure increases.
As the number of shells are increased using larger cutoffs, this is no longer true, due to the fact that the distance between the layers of fcc and hcp can be different depending on the pressure, thus the neighbour shells will no longer be isotropic.
Finally, with increasing cutoff, the energy contribution of the outmost shells gets smaller, the ``waves'' gradually flatten out.

Figure~\ref{fig:LJ_pd} shows the phase diagram of the simple energy shifted Lennard-Jones function (see eq.~\ref{eq:lj_e_shift}).
At smaller cutoff values, only one phase is found to be stable other than fcc and hcp, the $\langle \mathrm{hc} \rangle$ phase.
As the cutoff increases, first the $\langle \mathrm{hhc} \rangle$ and $\langle \mathrm{hhcc} \rangle$ phases appear, than other sequences with longer repeated subunits as well.

Applying additional shifts to the potential, such as force-shift and second-derivative shift, the ``wave'' like pattern of the phase diagram becomes significantly less pronounced (see the first two panels of Figure~\ref{fig:LJ_Fshift_pd}), but the order in which the more complex polytypes appear on the phase diagram remains similar.
For example $\langle \mathrm{hhc} \rangle$ and $\langle \mathrm{hhcc} \rangle$ phases appear on the second ``wave", the same two and $\langle \mathrm{hc} \rangle$ on the two sides of the third ``wave" and then $\langle \mathrm{hcccc} \rangle$ first appears on the tip of the fourth ``wave" in all three phase diagrams.
Although we are unable to offer a rigorous explanation for the flattening trend of the ``waves", we speculate that it is due to the fact that a non-smooth cut-off mechanism leads to large variations in energy as new neighbour shells cross the interaction range.
Since different polytypes have different neighbour shells, significant changes in energy differences can therefore occur.

Using a smoothing function to obtain a completely smooth potential, $U_{{\rm LJ}-C^\infty}$ has seemingly the opposite effect, while the width of the stability region of the $\langle \mathrm{hc} \rangle$ phase becomes significantly narrower, especially at shorter cutoffs, the magnitude of the ``waves'' increases (see Figure~\ref{fig:LJ_Fshift_pd}).
The explanation for this effect is that $U_{\rm LJ-C^\infty}$ is only {\em qualitatively smooth} but not {\em quantitively}.
Indeed we observe in Figure~\ref{fig:LJ_r_cutoff_types} that its second derivative has rapid variations in the interval where the cut-off is applied.
It also has to be noted that the exact effect of the smoothing will also depend on the widths of the smoothing region.

In order to study the effect of the Lennard-Jones exponents, thus the shape of the pair potential on the ground state phase diagram, we repeated our calculations on the force shifted potential with the following different $p$ and $q$ exponents: 12 and 8, 12 and 4, 14 and 6, 10 and 6. The phase diagrams are shown in Figure~\ref{fig:LJ_X_6_pd}.
It is clear from these figures that changing the exponents does not notably change the order in which the polytypic phases appear to be stable, thus the same stacking variants appear at the same cutoff values, but the corresponding pressure of the phase transitions are significantly different.
As a general rule, if either of the exponents are increased, the pressure above which the fcc phase is the most stable increases as well.

%\begin{figure}[hbt]
%\begin{center}
%\includegraphics[width=10cm]{LJ_12_x_Fshift_phase_diagram.pdf}
%\includegraphics[width=5cm, angle=90]{LJ12_8_Fshift_phase_diagram.pdf}
%\includegraphics[width=5cm, angle=90]{LJ_Fshift_phase_diagram.pdf}
%\includegraphics[width=5cm, angle=90]{LJ12_4_Fshift_phase_diagram.pdf}
%\end{center}
%\vspace{-40pt}
%\caption {Lennard-Jones $p=12$ and $q=8$ (top), $q=6$ (middle), $q=4$ (bottom) ground state structures as a function of cut-off radius (in sigma units) and pressure.
%The potential is force shifted.
%The colours represent the different stacking sequences, dark grey is pure fcc and light grey is pure hcp. }
%\label{fig:LJ_12_X_pd}
%\end{figure}

\section{Other interatomic potentials}

\subsection{Power law potential}

A simple repulsive power law potential with the exponent set to 12,
\begin{equation}
U_{\mathrm{pl}}(r) = 4\epsilon \Big(\frac{\sigma}{r}\Big)^{12},
\end{equation}
was also tested to see whether polytypic phases are stable in this case.
A force-shift was applied here as well.
The results indicate that the ground state structure is fcc at every pressure and cutoff studied.

\subsection{Morse potential}

\begin{figure}[hbt]
\begin{center}
\includegraphics[width=6.5cm, angle=90]{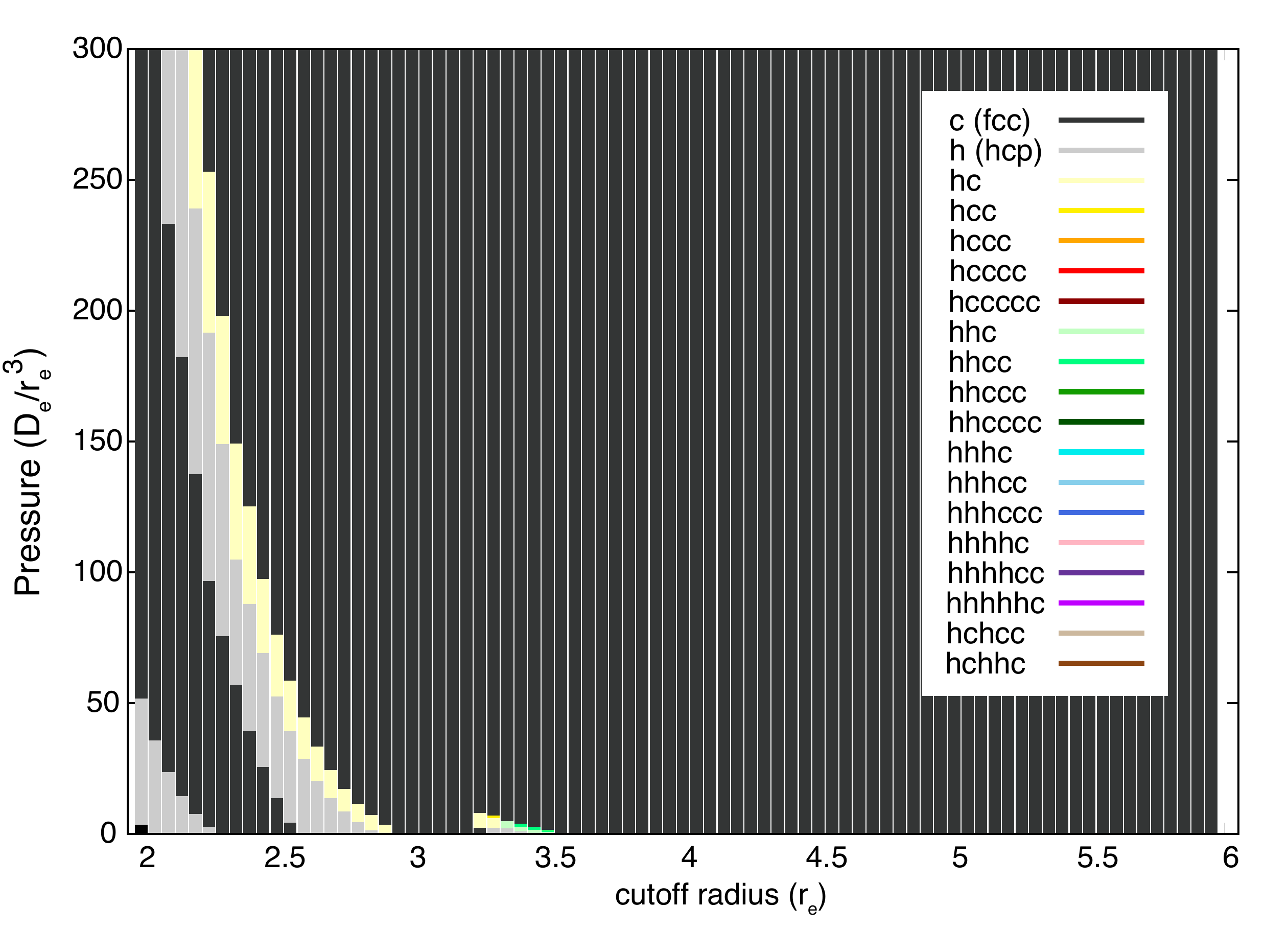}
\end{center}
\vspace{-15pt}
\caption {Morse potential ground state structures as a function of cut-off radius (in $r_e$ units) and pressure.
The potential is force shifted.
The colours represent the different stacking sequences, dark grey is pure fcc and light grey is pure hcp.
(Note: the pressure axis is different to that of the LJ one to show the low pressure region in more details.) }
\label{fig:Morse_pd}
\end{figure}

In order to see how another simple pair potential with an attractive term behave we also tested the Morse potential,
\begin{equation}
U_{\mathrm{M}}(r) = D_e \Big(e^{-2a(r-r_e)}-2e^{-a(r-r_e)}  \Big),
\end{equation}
with parameters $D_e=1.0$, $r_e=1.0$ controlling the depth and location of the minimum, respectively, and $a=4.0$. (We chose the value for parameter $a$ so that the pair potential is similar in shape to the Lennard-Jones potential.)
The potential was force shifted.
The Morse phase diagram (see Figure~\ref{fig:Morse_pd}) shows some stacking fault structures too: $\langle \mathrm{hc} \rangle$, $\langle \mathrm{hhc} \rangle$ and $\langle \mathrm{hhcc} \rangle$, similarly to the LJ potential, but only at small cutoffs.
Above $r_{cut}=3.6 r_e$ the fcc is the only stable phase.
This fast decay of the ``waves" can be explained by the exponential decay of the potential function, but the fact that polytypes other than fcc and hcp are found to be ground states also in case of the Morse potential indicates that the results we obtained for Lennard-Jones potentials in the previous sections might be generic for truncated pair potentials.

\section{Conclusion}

We have systematically studied the global minimum structure of the bulk Lennard-Jones model as a function of pressure and the details of potential truncation.
Our results demonstrate that its ground state structures are far more complex than previously reported, the stable phases including not only fcc and hcp but a wide range of more complex stacking sequences.
Most notably we obtained $\langle \mathrm{hc} \rangle$ and $\langle \mathrm{hhc} \rangle$ phases, the two polytypes most often observed in real materials (as dhcp and 9R) other then hcp and fcc.
This suggests that well-known pair potentials might be useful models of polytypism and can help us to understand and predict polytypic behaviour.

The relative stability of polytypes was found to be especially sensitive to the degree of smoothness of the potential around the cutoff.
This shows that the effect of truncation and the way the derivative of the potential is treated should not be underestimated when using pair potentials.
Further work is still needed, however, to obtain a clear theoretical explanation for the polytypism that we observed, e.g. to construct potentials with prescribed polytypes, as well as to confirm analogous effects in case of more complex model systems.

\begin{acknowledgements}
LBP acknowledges support from the Royal Society through a Dorothy Hodgkin Research Fellowship.
CO was supported by ERC Starting Grant 335120.
ABP was supported by a Leverhulme Early Career Fellowship and the Isaac Newton Trust until 2016.
CJP is also supported by the Royal Society through a Royal Society Wolfson Research Merit award.

%\texttt{https://www.repository.cam.ac.uk/handle/\\1810/255091}.
\end{acknowledgements}

\clearpage

\appendix

%\section{More on ANNNI}

%Due to the fact that polytypes are composed of virtually identical modular units, it is to be expected that their free energy will differ less than for polymorphs of other kinds.
%As a consequence it is frequently difficult to establish that the system is in true thermodynamic equilibrium.
%There are several theories to explain the polytypic behaviour, (i) crystal growth from a screw dislocation~\cite{Frank},with this any kind of polytype can be generated but its predictive power is limited and growth spirals are rarely seen in long-period polytypes (ii) polytypes are stabilised by vibrational entropy~\cite{Jagodzinski},
%but this has been since shown to be  an effect too small to contribute significantly~\cite{Weltner} (iii) energy of interaction of different units.

\section{Stable structures of the ANNNI and A3NNI models}

One of the simplest nontrivial models exhibiting periodically ordered phases, is the axial next-nearest-neighbour Ising model, ANNNI~\cite{Elliot_ANNNI,ANNNI_phaseD_Selke}, which is a variant of the Ising model with a two-state spin on each lattice site.
Interactions are between nearest neighbours, together with a second-neighbour interaction along one lattice direction (this is the axial direction, $z$).
The model is defined by the Hamiltonian
\begin{equation}
H = -\frac{1}{2}J_0 \sum_{ijj'}{S_{i,j}S_{i,j'}} - J_1 \sum_{ij}{S_{i,j}S_{i+1,j}} - J_2 \sum_{ij}{S_{i,j}S_{i+2,j}}
\end{equation}
where $S_{ij}$ is the two-state spin on each lattice site, and $i$ denotes the layers perpendicular to the axial direction and $j$ and $j'$ are nearest neighbour spins within the layer.

The ANNNI model is considered a prototype for polytypism~\cite{Price:1984vo, YEOMANS_Price1}, since its phase diagram contains sequences of long-wavelength-modulated phases, hence showing that short-range competing interactions are sufficient to stabilise long periodic structures as ground states.

Within ANNNI, the layers building up the polytypes can be characterised by two signs, $\downarrow$ and $\uparrow$.
In these models $\cdots\uparrow\uparrow\uparrow\cdots$ and $\cdots\downarrow\downarrow\downarrow\cdots$ are identical, and often simply marked as $\langle \infty \rangle$ (this is called the Zhdanov notation where the numbers in the brackets show the band widths, i.e. the number of layers with the same spin, e.g. $\langle 2 \rangle$=(2,2) means $\cdots\uparrow\uparrow\downarrow\downarrow\cdots$).

\begin{figure}[hbt]
\begin{center}
\includegraphics[width=6cm]{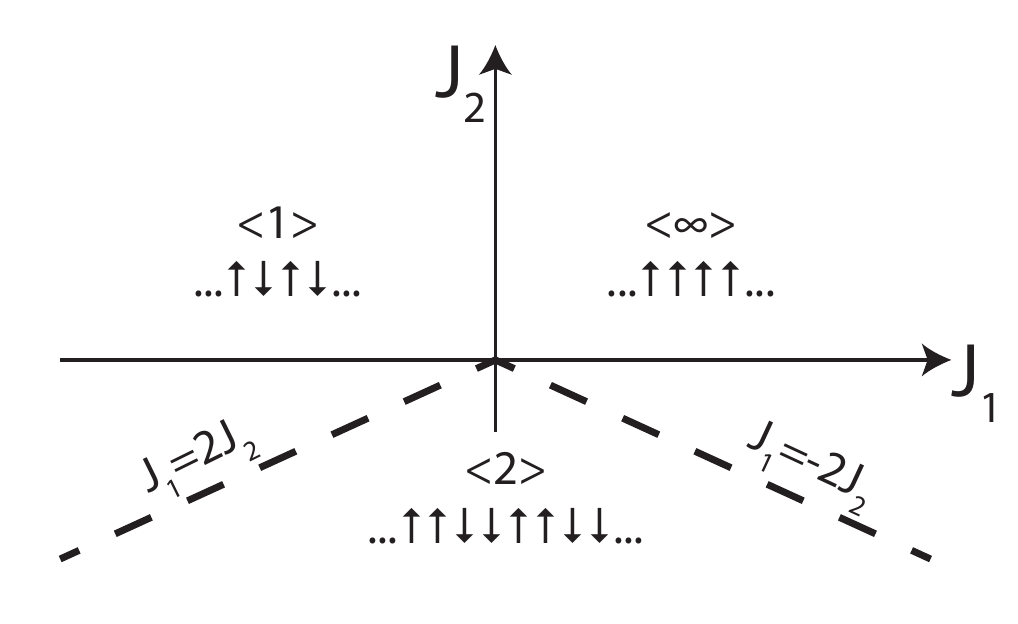}
\end{center}
\vspace{-10pt}
\caption {Ground states of the ANNNI model. Dashed lines mark multiphase boundaries.}
\label{fig:ANNNI_gs}
\end{figure}

The ground state phase diagram of ANNNI can be easily determined~ \cite{Price:1984vo}, see Figure~\ref{fig:ANNNI_gs}.
There are three main stable phases at 0~K, $\langle 1 \rangle$, $\langle 2 \rangle$ and $\langle \infty \rangle$, but the two dashed lines mark regions where the ground state is highly degenerate: along the line between $\langle 1 \rangle$ and $\langle 2 \rangle$ all phases containing only 1 and 2 bands have equal energy (e.g. $\langle 12 \rangle$,  $\langle 12^2 \rangle$,...etc.), and along the line between $\langle 2 \rangle$ and $\langle \infty \rangle$ all phases which contain no 1-bands have the same energy (e.g. $\langle 23 \rangle$,  $\langle 2^24 \rangle$, {\it etc.}).
Note, that the boundary between $\langle 1 \rangle$ and $\langle \infty \rangle$ is not degenerate! This means in particular that there are several phases missing from this phase diagram, e.g. $\langle 13 \rangle$, $\langle 14 \rangle$, and so forth, are not ground states at any value of $J_1$ or $J_2$.

\begin{figure}[htb]
\begin{center}
\includegraphics[width=8cm]{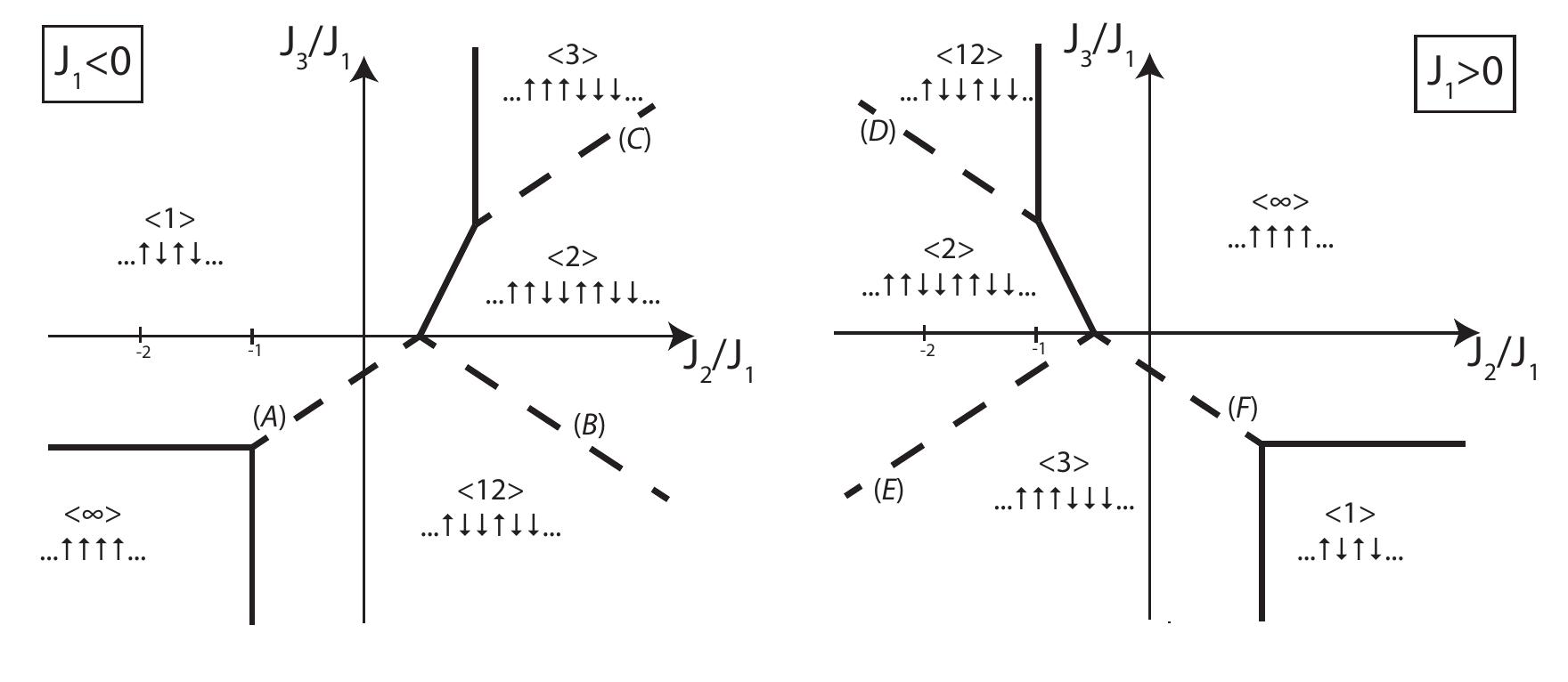}
\end{center}
\vspace{-10pt}
\caption {Ground states of the A3NNI model, in case $J_1 < 0$ (left hand side) and $J_1 > 0$ (right hand side).
Dashed lines mark multiphase boundaries and correspond to the following structure groups; $(A)$: $\langle 2(12)^k \rangle$ and $\langle 2(12)^k2(12)^{k+1} \rangle$ for $J_2/J_1<0$ and for $J_2/J_1>0$ also $\langle (12)^k112(12)^{k-1} \rangle$, $(B)$: $\langle 2(12)^k \rangle$ and $\langle 2(12)^k2(12)^{k+1} \rangle$, $(D)$: phases containing 1 and 2 bands, $(E)$: $\langle 23^k \rangle$ and $\langle 23^k23^{k+1} \rangle$, $(F)$: $\langle 3^k4 \rangle$ and for $J_2>0 \langle 3^k43^{k+1}4 \rangle$}
\label{fig:A3NNI_gs}
\end{figure}

In order to see how longer range interactions effect the stability of phases, the third neighbour Ising model, A3NNI has been studied too and discussed as a model for polytypism~\cite{YEOMANS:1988tl, MURAOKA1998773, Salje:1987wr}.
If the third neighbour interactions are also taken into account, the ground state phase diagram becomes more complicated and two additional structures appear as possible ground states phases compared to ANNNI, the $\langle12\rangle$ and $\langle3\rangle$ with several additional phases along the multiphase lines~\cite{YEOMANS:1988tl, Salje:1987wr}; see Figure~\ref{fig:A3NNI_gs}.

Many of the polytypic structures found in different materials resemble the phases seen in the ANNNI and A3NNI models.
For example, PbI$_2$ has a reversible phase transition between the phases $\langle \infty \rangle$ and $\langle 1 \rangle$, with the phases $\langle 2 \rangle$ and $\langle 12 \rangle$ being observed under different growth conditions only~\cite{Salje:1987wr}.
Same is true for ZnS and AgI.
Phase transitions between $\langle 1 \rangle - \langle 2 \rangle - \langle \infty \rangle$ are found in MgSiO$_3$, and $\langle 1 \rangle - \langle \infty \rangle - \langle 3 \rangle - \langle 23 \rangle $ are seen in case of SiC~\cite{Jepps}.
However, there are also polytypic structures seen in experiments, e.g. $\langle 13 \rangle$ in spinelloids, which do not occur in the ANNNI model.

At temperatures slightly above 0~K, the phase diagram remains similar but at the vicinity of the multiphase lines sequences of $\langle12^k\rangle$ and $\langle2^k3\rangle$ appear, and even more new phases at higher temperatures (though the proportion of disordered layers increase too)~\cite{ANNNI_phaseD_Selke}.

\bibliography{LJ_sf_references}

%merlin.mbs apsrev4-1.bst 2010-07-25 4.21a (PWD, AO, DPC) hacked
%Control: key (0)
%Control: author (8) initials jnrlst
%Control: editor formatted (1) identically to author
%Control: production of article title (-1) disabled
%Control: page (0) single
%Control: year (1) truncated
%Control: production of eprint (0) enabled
\begin{thebibliography}{38}%
\makeatletter
\providecommand \@ifxundefined [1]{%
 \@ifx{#1\undefined}
}%
\providecommand \@ifnum [1]{%
 \ifnum #1\expandafter \@firstoftwo
 \else \expandafter \@secondoftwo
 \fi
}%
\providecommand \@ifx [1]{%
 \ifx #1\expandafter \@firstoftwo
 \else \expandafter \@secondoftwo
 \fi
}%
\providecommand \natexlab [1]{#1}%
\providecommand \enquote  [1]{``#1''}%
\providecommand \bibnamefont  [1]{#1}%
\providecommand \bibfnamefont [1]{#1}%
\providecommand \citenamefont [1]{#1}%
\providecommand \href@noop [0]{\@secondoftwo}%
\providecommand \href [0]{\begingroup \@sanitize@url \@href}%
\providecommand \@href[1]{\@@startlink{#1}\@@href}%
\providecommand \@@href[1]{\endgroup#1\@@endlink}%
\providecommand \@sanitize@url [0]{\catcode `\\12\catcode `\$12\catcode
  `\&12\catcode `\#12\catcode `\^12\catcode `\_12\catcode `\%12\relax}%
\providecommand \@@startlink[1]{}%
\providecommand \@@endlink[0]{}%
\providecommand \url  [0]{\begingroup\@sanitize@url \@url }%
\providecommand \@url [1]{\endgroup\@href {#1}{\urlprefix }}%
\providecommand \urlprefix  [0]{URL }%
\providecommand \Eprint [0]{\href }%
\providecommand \doibase [0]{http://dx.doi.org/}%
\providecommand \selectlanguage [0]{\@gobble}%
\providecommand \bibinfo  [0]{\@secondoftwo}%
\providecommand \bibfield  [0]{\@secondoftwo}%
\providecommand \translation [1]{[#1]}%
\providecommand \BibitemOpen [0]{}%
\providecommand \bibitemStop [0]{}%
\providecommand \bibitemNoStop [0]{.\EOS\space}%
\providecommand \EOS [0]{\spacefactor3000\relax}%
\providecommand \BibitemShut  [1]{\csname bibitem#1\endcsname}%
\let\auto@bib@innerbib\@empty
%</preamble>
\bibitem [{\citenamefont {Trigunayat}(1991)}]{poly_review}%
  \BibitemOpen
  \bibfield  {author} {\bibinfo {author} {\bibfnamefont {G.~C.}\ \bibnamefont
  {Trigunayat}},\ }\href@noop {} {\bibfield  {journal} {\bibinfo  {journal}
  {Solid State Ionics}\ }\textbf {\bibinfo {volume} {48}},\ \bibinfo {pages}
  {3} (\bibinfo {year} {1991})}\BibitemShut {NoStop}%
\bibitem [{\citenamefont {Jepps}\ and\ \citenamefont {Page}(1983)}]{Jepps}%
  \BibitemOpen
  \bibfield  {author} {\bibinfo {author} {\bibfnamefont {N.~W.}\ \bibnamefont
  {Jepps}}\ and\ \bibinfo {author} {\bibfnamefont {T.~F.}\ \bibnamefont
  {Page}},\ }\href@noop {} {\bibfield  {journal} {\bibinfo  {journal} {J.
  Cryst. Growth Charact.}\ }\textbf {\bibinfo {volume} {7}},\ \bibinfo {pages}
  {259 } (\bibinfo {year} {1983})}\BibitemShut {NoStop}%
\bibitem [{\citenamefont {Nakashima}\ and\ \citenamefont
  {Hangyo}(1991)}]{SiC_raman}%
  \BibitemOpen
  \bibfield  {author} {\bibinfo {author} {\bibfnamefont {S.}~\bibnamefont
  {Nakashima}}\ and\ \bibinfo {author} {\bibfnamefont {M.}~\bibnamefont
  {Hangyo}},\ }\href@noop {} {\bibfield  {journal} {\bibinfo  {journal} {Solid
  State Comm.}\ }\textbf {\bibinfo {volume} {80}},\ \bibinfo {pages} {21}
  (\bibinfo {year} {1991})}\BibitemShut {NoStop}%
\bibitem [{\citenamefont {Mardix}\ \emph {et~al.}(1969)\citenamefont {Mardix},
  \citenamefont {Kiflawi},\ and\ \citenamefont {Kalman}}]{ZnS}%
  \BibitemOpen
  \bibfield  {author} {\bibinfo {author} {\bibfnamefont {S.}~\bibnamefont
  {Mardix}}, \bibinfo {author} {\bibfnamefont {I.}~\bibnamefont {Kiflawi}}, \
  and\ \bibinfo {author} {\bibfnamefont {Z.~H.}\ \bibnamefont {Kalman}},\
  }\href@noop {} {\bibfield  {journal} {\bibinfo  {journal} {Acta Crystallogr.
  B}\ }\textbf {\bibinfo {volume} {B25}},\ \bibinfo {pages} {1586} (\bibinfo
  {year} {1969})}\BibitemShut {NoStop}%
\bibitem [{\citenamefont {Boutaiba}\ \emph {et~al.}(2014)\citenamefont
  {Boutaiba}, \citenamefont {Belabbes}, \citenamefont {Ferhat},\ and\
  \citenamefont {Bechstedt}}]{ZnS_theory}%
  \BibitemOpen
  \bibfield  {author} {\bibinfo {author} {\bibfnamefont {F.}~\bibnamefont
  {Boutaiba}}, \bibinfo {author} {\bibfnamefont {A.}~\bibnamefont {Belabbes}},
  \bibinfo {author} {\bibfnamefont {M.}~\bibnamefont {Ferhat}}, \ and\ \bibinfo
  {author} {\bibfnamefont {F.}~\bibnamefont {Bechstedt}},\ }\href@noop {}
  {\bibfield  {journal} {\bibinfo  {journal} {Phys. Rev. B}\ }\textbf {\bibinfo
  {volume} {89}},\ \bibinfo {pages} {245308} (\bibinfo {year}
  {2014})}\BibitemShut {NoStop}%
\bibitem [{\citenamefont {Katahama}\ \emph {et~al.}(1984)\citenamefont
  {Katahama}, \citenamefont {Nakashima}, \citenamefont {Hangyo}, \citenamefont
  {Mitsuishi},\ and\ \citenamefont {Palosz}}]{CdI2}%
  \BibitemOpen
  \bibfield  {author} {\bibinfo {author} {\bibfnamefont {H.}~\bibnamefont
  {Katahama}}, \bibinfo {author} {\bibfnamefont {S.}~\bibnamefont {Nakashima}},
  \bibinfo {author} {\bibfnamefont {M.}~\bibnamefont {Hangyo}}, \bibinfo
  {author} {\bibfnamefont {A.}~\bibnamefont {Mitsuishi}}, \ and\ \bibinfo
  {author} {\bibfnamefont {B.}~\bibnamefont {Palosz}},\ }\href@noop {}
  {\bibfield  {journal} {\bibinfo  {journal} {Solid State Commun.}\ }\textbf
  {\bibinfo {volume} {49}},\ \bibinfo {pages} {547} (\bibinfo {year}
  {1984})}\BibitemShut {NoStop}%
\bibitem [{\citenamefont {Zagorac}\ \emph {et~al.}(2015)\citenamefont
  {Zagorac}, \citenamefont {Sch\"{o}n}, \citenamefont {Zagoracab},\ and\
  \citenamefont {Jansena}}]{ZnO}%
  \BibitemOpen
  \bibfield  {author} {\bibinfo {author} {\bibfnamefont {D.}~\bibnamefont
  {Zagorac}}, \bibinfo {author} {\bibfnamefont {J.~C.}\ \bibnamefont
  {Sch\"{o}n}}, \bibinfo {author} {\bibfnamefont {J.}~\bibnamefont
  {Zagoracab}}, \ and\ \bibinfo {author} {\bibfnamefont {M.}~\bibnamefont
  {Jansena}},\ }\href@noop {} {\bibfield  {journal} {\bibinfo  {journal} {RSC
  Advances}\ }\textbf {\bibinfo {volume} {5}},\ \bibinfo {pages} {25929}
  (\bibinfo {year} {2015})}\BibitemShut {NoStop}%
\bibitem [{\citenamefont {Wen}\ \emph {et~al.}(2008)\citenamefont {Wen},
  \citenamefont {Zhao}, \citenamefont {Bucknum}, \citenamefont {Yao},\ and\
  \citenamefont {Li}}]{diamond}%
  \BibitemOpen
  \bibfield  {author} {\bibinfo {author} {\bibfnamefont {B.}~\bibnamefont
  {Wen}}, \bibinfo {author} {\bibfnamefont {J.}~\bibnamefont {Zhao}}, \bibinfo
  {author} {\bibfnamefont {M.~J.}\ \bibnamefont {Bucknum}}, \bibinfo {author}
  {\bibfnamefont {P.}~\bibnamefont {Yao}}, \ and\ \bibinfo {author}
  {\bibfnamefont {T.}~\bibnamefont {Li}},\ }\href@noop {} {\bibfield  {journal}
  {\bibinfo  {journal} {Diam. Relat. Mater.}\ }\textbf {\bibinfo {volume}
  {17}},\ \bibinfo {pages} {356} (\bibinfo {year} {2008})}\BibitemShut
  {NoStop}%
\bibitem [{\citenamefont {Salzmann}\ \emph {et~al.}(2015)\citenamefont
  {Salzmann}, \citenamefont {Murray},\ and\ \citenamefont
  {Shephard}}]{diamond_xray}%
  \BibitemOpen
  \bibfield  {author} {\bibinfo {author} {\bibfnamefont {C.~G.}\ \bibnamefont
  {Salzmann}}, \bibinfo {author} {\bibfnamefont {B.~J.}\ \bibnamefont
  {Murray}}, \ and\ \bibinfo {author} {\bibfnamefont {J.~J.}\ \bibnamefont
  {Shephard}},\ }\href@noop {} {\bibfield  {journal} {\bibinfo  {journal}
  {Diam. Relat. Mater.}\ }\textbf {\bibinfo {volume} {59}},\ \bibinfo {pages}
  {69} (\bibinfo {year} {2015})}\BibitemShut {NoStop}%
\bibitem [{\citenamefont {Ba\u{g}ci}\ \emph {et~al.}(2010)\citenamefont
  {Ba\u{g}ci}, \citenamefont {T\"{u}t\"{u}nc\"{u}}, \citenamefont {Duman},\
  and\ \citenamefont {Srivastava}}]{La_dhcp}%
  \BibitemOpen
  \bibfield  {author} {\bibinfo {author} {\bibfnamefont {S.}~\bibnamefont
  {Ba\u{g}ci}}, \bibinfo {author} {\bibfnamefont {H.~M.}\ \bibnamefont
  {T\"{u}t\"{u}nc\"{u}}}, \bibinfo {author} {\bibfnamefont {S.}~\bibnamefont
  {Duman}}, \ and\ \bibinfo {author} {\bibfnamefont {G.~P.}\ \bibnamefont
  {Srivastava}},\ }\href@noop {} {\bibfield  {journal} {\bibinfo  {journal}
  {Phys. Rev. B.}\ }\textbf {\bibinfo {volume} {81}},\ \bibinfo {pages}
  {144507} (\bibinfo {year} {2010})}\BibitemShut {NoStop}%
\bibitem [{\citenamefont {Ellinger}\ and\ \citenamefont
  {Zachariasen}(1953)}]{Sm_9R}%
  \BibitemOpen
  \bibfield  {author} {\bibinfo {author} {\bibfnamefont {F.~H.}\ \bibnamefont
  {Ellinger}}\ and\ \bibinfo {author} {\bibfnamefont {W.~H.}\ \bibnamefont
  {Zachariasen}},\ }\href@noop {} {\bibfield  {journal} {\bibinfo  {journal}
  {J. Am. Chem. Soc.}\ }\textbf {\bibinfo {volume} {75}},\ \bibinfo {pages}
  {5650} (\bibinfo {year} {1953})}\BibitemShut {NoStop}%
\bibitem [{\citenamefont {Overhauser}(1984)}]{Li_9R}%
  \BibitemOpen
  \bibfield  {author} {\bibinfo {author} {\bibfnamefont {A.~W.}\ \bibnamefont
  {Overhauser}},\ }\href@noop {} {\bibfield  {journal} {\bibinfo  {journal}
  {Phys. Rev. Lett.}\ }\textbf {\bibinfo {volume} {53}},\ \bibinfo {pages} {64}
  (\bibinfo {year} {1984})}\BibitemShut {NoStop}%
\bibitem [{\citenamefont {Samudrala}\ \emph {et~al.}(2011)\citenamefont
  {Samudrala}, \citenamefont {Thomas}, \citenamefont {Montgomery},\ and\
  \citenamefont {Vohra}}]{Er_dhcp_9R}%
  \BibitemOpen
  \bibfield  {author} {\bibinfo {author} {\bibfnamefont {G.~K.}\ \bibnamefont
  {Samudrala}}, \bibinfo {author} {\bibfnamefont {S.~A.}\ \bibnamefont
  {Thomas}}, \bibinfo {author} {\bibfnamefont {J.~M.}\ \bibnamefont
  {Montgomery}}, \ and\ \bibinfo {author} {\bibfnamefont {Y.~K.}\ \bibnamefont
  {Vohra}},\ }\href@noop {} {\bibfield  {journal} {\bibinfo  {journal} {J.
  Phys. Condens. Matter}\ }\textbf {\bibinfo {volume} {23}},\ \bibinfo {pages}
  {315701} (\bibinfo {year} {2011})}\BibitemShut {NoStop}%
\bibitem [{\citenamefont {Shu}\ \emph {et~al.}(2016)\citenamefont {Shu},
  \citenamefont {Hu}, \citenamefont {Liu}, \citenamefont {Shen}, \citenamefont
  {Xu}, \citenamefont {Zhao}, \citenamefont {He}, \citenamefont {Wang},
  \citenamefont {Tian},\ and\ \citenamefont {Yu}}]{Bi_poly}%
  \BibitemOpen
  \bibfield  {author} {\bibinfo {author} {\bibfnamefont {Y.}~\bibnamefont
  {Shu}}, \bibinfo {author} {\bibfnamefont {W.}~\bibnamefont {Hu}}, \bibinfo
  {author} {\bibfnamefont {Z.}~\bibnamefont {Liu}}, \bibinfo {author}
  {\bibfnamefont {G.}~\bibnamefont {Shen}}, \bibinfo {author} {\bibfnamefont
  {B.}~\bibnamefont {Xu}}, \bibinfo {author} {\bibfnamefont {Z.}~\bibnamefont
  {Zhao}}, \bibinfo {author} {\bibfnamefont {J.}~\bibnamefont {He}}, \bibinfo
  {author} {\bibfnamefont {Y.}~\bibnamefont {Wang}}, \bibinfo {author}
  {\bibfnamefont {Y.}~\bibnamefont {Tian}}, \ and\ \bibinfo {author}
  {\bibfnamefont {D.}~\bibnamefont {Yu}},\ }\href@noop {} {\bibfield  {journal}
  {\bibinfo  {journal} {Sci. Rep.}\ }\textbf {\bibinfo {volume} {6}},\ \bibinfo
  {pages} {20337} (\bibinfo {year} {2016})}\BibitemShut {NoStop}%
\bibitem [{\citenamefont {Jephcoat}\ \emph {et~al.}(1987)\citenamefont
  {Jephcoat}, \citenamefont {Mao}, \citenamefont {Finger}, \citenamefont {Cox},
  \citenamefont {Hemley},\ and\ \citenamefont {Zha}}]{Jephcoat_KrXe}%
  \BibitemOpen
  \bibfield  {author} {\bibinfo {author} {\bibfnamefont {A.~P.}\ \bibnamefont
  {Jephcoat}}, \bibinfo {author} {\bibfnamefont {H.~K.}\ \bibnamefont {Mao}},
  \bibinfo {author} {\bibfnamefont {L.~W.}\ \bibnamefont {Finger}}, \bibinfo
  {author} {\bibfnamefont {D.}~\bibnamefont {Cox}}, \bibinfo {author}
  {\bibfnamefont {R.~J.}\ \bibnamefont {Hemley}}, \ and\ \bibinfo {author}
  {\bibfnamefont {C.~S.}\ \bibnamefont {Zha}},\ }\href@noop {} {\bibfield
  {journal} {\bibinfo  {journal} {Phys. Rev. Lett.}\ }\textbf {\bibinfo
  {volume} {59}},\ \bibinfo {pages} {2670} (\bibinfo {year}
  {1987})}\BibitemShut {NoStop}%
\bibitem [{\citenamefont {Errandonea}\ \emph {et~al.}(2002)\citenamefont
  {Errandonea}, \citenamefont {Schwager}, \citenamefont {Boehler},\ and\
  \citenamefont {Ross}}]{Errandonea:2002hx}%
  \BibitemOpen
  \bibfield  {author} {\bibinfo {author} {\bibfnamefont {D.}~\bibnamefont
  {Errandonea}}, \bibinfo {author} {\bibfnamefont {B.}~\bibnamefont
  {Schwager}}, \bibinfo {author} {\bibfnamefont {R.}~\bibnamefont {Boehler}}, \
  and\ \bibinfo {author} {\bibfnamefont {M.}~\bibnamefont {Ross}},\ }\href@noop
  {} {\bibfield  {journal} {\bibinfo  {journal} {Phys. Rev. B}\ }\textbf
  {\bibinfo {volume} {65}},\ \bibinfo {pages} {214110} (\bibinfo {year}
  {2002})}\BibitemShut {NoStop}%
\bibitem [{\citenamefont {Yoo}\ \emph {et~al.}(1996)\citenamefont {Yoo},
  \citenamefont {S\"{o}derlind}, \citenamefont {Moriarty},\ and\ \citenamefont
  {Cambell}}]{dhcp_Fe1}%
  \BibitemOpen
  \bibfield  {author} {\bibinfo {author} {\bibfnamefont {C.}~\bibnamefont
  {Yoo}}, \bibinfo {author} {\bibfnamefont {P.}~\bibnamefont {S\"{o}derlind}},
  \bibinfo {author} {\bibfnamefont {J.}~\bibnamefont {Moriarty}}, \ and\
  \bibinfo {author} {\bibfnamefont {A.}~\bibnamefont {Cambell}},\ }\href@noop
  {} {\bibfield  {journal} {\bibinfo  {journal} {Phys. Lett. A}\ }\textbf
  {\bibinfo {volume} {214}},\ \bibinfo {pages} {65} (\bibinfo {year}
  {1996})}\BibitemShut {NoStop}%
\bibitem [{\citenamefont {Mikhaylushkin}\ \emph {et~al.}(2007)\citenamefont
  {Mikhaylushkin}, \citenamefont {Simak}, \citenamefont {Dubrovinsky},
  \citenamefont {Dubrovinskaia}, \citenamefont {Johansson},\ and\ \citenamefont
  {Abrikosov}}]{dhcp_Fe2}%
  \BibitemOpen
  \bibfield  {author} {\bibinfo {author} {\bibfnamefont {A.~S.}\ \bibnamefont
  {Mikhaylushkin}}, \bibinfo {author} {\bibfnamefont {S.~I.}\ \bibnamefont
  {Simak}}, \bibinfo {author} {\bibfnamefont {L.}~\bibnamefont {Dubrovinsky}},
  \bibinfo {author} {\bibfnamefont {N.}~\bibnamefont {Dubrovinskaia}}, \bibinfo
  {author} {\bibfnamefont {B.}~\bibnamefont {Johansson}}, \ and\ \bibinfo
  {author} {\bibfnamefont {I.~A.}\ \bibnamefont {Abrikosov}},\ }\href@noop {}
  {\bibfield  {journal} {\bibinfo  {journal} {Phys. Rev. Lett.}\ }\textbf
  {\bibinfo {volume} {99}},\ \bibinfo {pages} {165505} (\bibinfo {year}
  {2007})}\BibitemShut {NoStop}%
\bibitem [{\citenamefont {Price}\ and\ \citenamefont
  {Yeomans}(1984)}]{Price:1984vo}%
  \BibitemOpen
  \bibfield  {author} {\bibinfo {author} {\bibfnamefont {G.~D.}\ \bibnamefont
  {Price}}\ and\ \bibinfo {author} {\bibfnamefont {J.}~\bibnamefont
  {Yeomans}},\ }\href@noop {} {\bibfield  {journal} {\bibinfo  {journal} {Acta
  Crystallogr. B}\ }\textbf {\bibinfo {volume} {40}},\ \bibinfo {pages} {448}
  (\bibinfo {year} {1984})}\BibitemShut {NoStop}%
\bibitem [{\citenamefont {{J Yeomans and G D Price}}(1986)}]{YEOMANS_Price1}%
  \BibitemOpen
  \bibfield  {author} {\bibinfo {author} {\bibnamefont {{J Yeomans and G D
  Price}}},\ }\href@noop {} {\bibfield  {journal} {\bibinfo  {journal} {B.
  Min\'eral.}\ }\textbf {\bibinfo {volume} {109}},\ \bibinfo {pages} {3}
  (\bibinfo {year} {1986})}\BibitemShut {NoStop}%
\bibitem [{\citenamefont {{J Yeomans}}(1988)}]{YEOMANS:1988tl}%
  \BibitemOpen
  \bibfield  {author} {\bibinfo {author} {\bibnamefont {{J Yeomans}}},\
  }\href@noop {} {\bibfield  {journal} {\bibinfo  {journal} {Solid State
  Phys.}\ }\textbf {\bibinfo {volume} {41}},\ \bibinfo {pages} {151} (\bibinfo
  {year} {1988})}\BibitemShut {NoStop}%
\bibitem [{\citenamefont {Kihara}\ and\ \citenamefont
  {Koba}(1952)}]{Koba_hcpfcc_52}%
  \BibitemOpen
  \bibfield  {author} {\bibinfo {author} {\bibfnamefont {T.}~\bibnamefont
  {Kihara}}\ and\ \bibinfo {author} {\bibfnamefont {S.}~\bibnamefont {Koba}},\
  }\href@noop {} {\bibfield  {journal} {\bibinfo  {journal} {J. Phys. Soc.
  Jpn.}\ }\textbf {\bibinfo {volume} {7}},\ \bibinfo {pages} {348} (\bibinfo
  {year} {1952})}\BibitemShut {NoStop}%
\bibitem [{\citenamefont {Travesset}(2014)}]{DLT_LJ_crystal}%
  \BibitemOpen
  \bibfield  {author} {\bibinfo {author} {\bibfnamefont {A.}~\bibnamefont
  {Travesset}},\ }\href@noop {} {\bibfield  {journal} {\bibinfo  {journal} {The
  Journal of Chemical Physics}\ }\textbf {\bibinfo {volume} {141}} (\bibinfo
  {year} {2014})}\BibitemShut {NoStop}%
\bibitem [{\citenamefont {de~Souza}\ and\ \citenamefont
  {Wales}(2016)}]{Wales_LJ2016}%
  \BibitemOpen
  \bibfield  {author} {\bibinfo {author} {\bibfnamefont {V.~K.}\ \bibnamefont
  {de~Souza}}\ and\ \bibinfo {author} {\bibfnamefont {D.~J.}\ \bibnamefont
  {Wales}},\ }\href@noop {} {\bibfield  {journal} {\bibinfo  {journal} {Journal
  of Statistical Mechanics: Theory and Experiment}\ }\textbf {\bibinfo {volume}
  {7}},\ \bibinfo {pages} {074001} (\bibinfo {year} {2016})}\BibitemShut
  {NoStop}%
\bibitem [{\citenamefont {Smit}(1992)}]{BSmit_LJ_vap}%
  \BibitemOpen
  \bibfield  {author} {\bibinfo {author} {\bibfnamefont {B.}~\bibnamefont
  {Smit}},\ }\href@noop {} {\bibfield  {journal} {\bibinfo  {journal} {J. Chem.
  Phys}\ }\textbf {\bibinfo {volume} {96}},\ \bibinfo {pages} {8639} (\bibinfo
  {year} {1992})}\BibitemShut {NoStop}%
\bibitem [{\citenamefont {Panagiotopoulos}(1994)}]{Panagiotopoulos_LJ_vap}%
  \BibitemOpen
  \bibfield  {author} {\bibinfo {author} {\bibfnamefont {A.~Z.}\ \bibnamefont
  {Panagiotopoulos}},\ }\href@noop {} {\bibfield  {journal} {\bibinfo
  {journal} {Int. J. Thermophys.}\ }\textbf {\bibinfo {volume} {15}},\ \bibinfo
  {pages} {1057} (\bibinfo {year} {1994})}\BibitemShut {NoStop}%
\bibitem [{\citenamefont {Shi}\ and\ \citenamefont
  {Johnson}(2001)}]{Johnson_LJ_vap}%
  \BibitemOpen
  \bibfield  {author} {\bibinfo {author} {\bibfnamefont {W.}~\bibnamefont
  {Shi}}\ and\ \bibinfo {author} {\bibfnamefont {J.~K.}\ \bibnamefont
  {Johnson}},\ }\href@noop {} {\bibfield  {journal} {\bibinfo  {journal} {Fluid
  Phase Equilibr.}\ }\textbf {\bibinfo {volume} {187--188}},\ \bibinfo {pages}
  {171} (\bibinfo {year} {2001})}\BibitemShut {NoStop}%
\bibitem [{\citenamefont {Mastny}\ and\ \citenamefont
  {de~Pablo}(2007)}]{dePablo_LJ_melt}%
  \BibitemOpen
  \bibfield  {author} {\bibinfo {author} {\bibfnamefont {E.~A.}\ \bibnamefont
  {Mastny}}\ and\ \bibinfo {author} {\bibfnamefont {J.~J.}\ \bibnamefont
  {de~Pablo}},\ }\href@noop {} {\bibfield  {journal} {\bibinfo  {journal} {J.
  Chem. Phys}\ }\textbf {\bibinfo {volume} {127}},\ \bibinfo {pages} {104504}
  (\bibinfo {year} {2007})}\BibitemShut {NoStop}%
\bibitem [{\citenamefont {Ahmed}\ and\ \citenamefont
  {Sadusa}(2010)}]{Sadusa_LJ_melt}%
  \BibitemOpen
  \bibfield  {author} {\bibinfo {author} {\bibfnamefont {A.}~\bibnamefont
  {Ahmed}}\ and\ \bibinfo {author} {\bibfnamefont {R.~J.}\ \bibnamefont
  {Sadusa}},\ }\href@noop {} {\bibfield  {journal} {\bibinfo  {journal} {J.
  Chem. Phys}\ }\textbf {\bibinfo {volume} {133}},\ \bibinfo {pages} {124515}
  (\bibinfo {year} {2010})}\BibitemShut {NoStop}%
\bibitem [{\citenamefont {van~der Hoef}(2000)}]{LJ_fcc_Hoef}%
  \BibitemOpen
  \bibfield  {author} {\bibinfo {author} {\bibfnamefont {M.~A.}\ \bibnamefont
  {van~der Hoef}},\ }\href@noop {} {\bibfield  {journal} {\bibinfo  {journal}
  {J. Chem. Phys.}\ }\textbf {\bibinfo {volume} {113}},\ \bibinfo {pages}
  {8142} (\bibinfo {year} {2000})}\BibitemShut {NoStop}%
\bibitem [{\citenamefont {van~de Waal}(1991)}]{Waal_LJ}%
  \BibitemOpen
  \bibfield  {author} {\bibinfo {author} {\bibfnamefont {B.~W.}\ \bibnamefont
  {van~de Waal}},\ }\href@noop {} {\bibfield  {journal} {\bibinfo  {journal}
  {Phys. Rev. Lett.}\ }\textbf {\bibinfo {volume} {67}},\ \bibinfo {pages}
  {3263} (\bibinfo {year} {1991})}\BibitemShut {NoStop}%
\bibitem [{\citenamefont {Jackson}\ \emph {et~al.}(2002)\citenamefont
  {Jackson}, \citenamefont {Bruce},\ and\ \citenamefont
  {Ackland}}]{LJ_cutoffs_Ackland}%
  \BibitemOpen
  \bibfield  {author} {\bibinfo {author} {\bibfnamefont {A.~N.}\ \bibnamefont
  {Jackson}}, \bibinfo {author} {\bibfnamefont {A.~D.}\ \bibnamefont {Bruce}},
  \ and\ \bibinfo {author} {\bibfnamefont {G.~J.}\ \bibnamefont {Ackland}},\
  }\href@noop {} {\bibfield  {journal} {\bibinfo  {journal} {Phys. Rev. E}\
  }\textbf {\bibinfo {volume} {65}},\ \bibinfo {pages} {036710} (\bibinfo
  {year} {2002})}\BibitemShut {NoStop}%
\bibitem [{\citenamefont {Li}(2003)}]{Li:2003wy}%
  \BibitemOpen
  \bibfield  {author} {\bibinfo {author} {\bibfnamefont {J.}~\bibnamefont
  {Li}},\ }\href@noop {} {\bibfield  {journal} {\bibinfo  {journal} {Model.
  Simul. Mater. Sci. Eng.}\ }\textbf {\bibinfo {volume} {11}},\ \bibinfo
  {pages} {173} (\bibinfo {year} {2003})},\ \bibinfo {note} {we used the
  enhanced version of the AtomEye atomistic configuration viewer, provided by
  James Kermode at \url{http://www.jrkermode.co.uk}}\BibitemShut {NoStop}%
\bibitem [{qui()}]{quip}%
  \BibitemOpen
  \href@noop {} {\enquote {\bibinfo {title} {{libAtoms package}},}\ }\bibinfo
  {howpublished} {\url{http://www.libatoms.org}}\BibitemShut {NoStop}%
\bibitem [{\citenamefont {Elliott}(1961)}]{Elliot_ANNNI}%
  \BibitemOpen
  \bibfield  {author} {\bibinfo {author} {\bibfnamefont {R.~J.}\ \bibnamefont
  {Elliott}},\ }\href@noop {} {\bibfield  {journal} {\bibinfo  {journal} {Phys.
  Rev.}\ }\textbf {\bibinfo {volume} {124}},\ \bibinfo {pages} {346} (\bibinfo
  {year} {1961})}\BibitemShut {NoStop}%
\bibitem [{\citenamefont {Fisher}\ and\ \citenamefont
  {Selke}(1980)}]{ANNNI_phaseD_Selke}%
  \BibitemOpen
  \bibfield  {author} {\bibinfo {author} {\bibfnamefont {M.~E.}\ \bibnamefont
  {Fisher}}\ and\ \bibinfo {author} {\bibfnamefont {W.}~\bibnamefont {Selke}},\
  }\href@noop {} {\bibfield  {journal} {\bibinfo  {journal} {Phys. Rev. Lett.}\
  }\textbf {\bibinfo {volume} {44}},\ \bibinfo {pages} {1502} (\bibinfo {year}
  {1980})}\BibitemShut {NoStop}%
\bibitem [{\citenamefont {Muraoka}\ \emph {et~al.}(1998)\citenamefont
  {Muraoka}, \citenamefont {Kanemaru},\ and\ \citenamefont
  {Idogaki}}]{MURAOKA1998773}%
  \BibitemOpen
  \bibfield  {author} {\bibinfo {author} {\bibfnamefont {Y.}~\bibnamefont
  {Muraoka}}, \bibinfo {author} {\bibfnamefont {M.}~\bibnamefont {Kanemaru}}, \
  and\ \bibinfo {author} {\bibfnamefont {T.}~\bibnamefont {Idogaki}},\
  }\href@noop {} {\bibfield  {journal} {\bibinfo  {journal} {J. Magn. Magn.
  Mater.}\ }\textbf {\bibinfo {volume} {177}},\ \bibinfo {pages} {773 }
  (\bibinfo {year} {1998})}\BibitemShut {NoStop}%
\bibitem [{\citenamefont {Salje}\ \emph {et~al.}(1987)\citenamefont {Salje},
  \citenamefont {Palosz},\ and\ \citenamefont {Wruck}}]{Salje:1987wr}%
  \BibitemOpen
  \bibfield  {author} {\bibinfo {author} {\bibfnamefont {E.}~\bibnamefont
  {Salje}}, \bibinfo {author} {\bibfnamefont {B.}~\bibnamefont {Palosz}}, \
  and\ \bibinfo {author} {\bibfnamefont {B.}~\bibnamefont {Wruck}},\
  }\href@noop {} {\bibfield  {journal} {\bibinfo  {journal} {J. Phys. C Solid
  State}\ }\textbf {\bibinfo {volume} {20}},\ \bibinfo {pages} {4077} (\bibinfo
  {year} {1987})}\BibitemShut {NoStop}%
\end{thebibliography}%

\end{document}